\documentclass[%
 reprint,
 amsmath,amssymb,
 aps,
]{revtex4-2}

\usepackage{pkgfile}
\usepackage{xifthen}
\usepackage{tabularx}
\usepackage{adjustbox}
\usepackage{bbm}
\usepackage{makecell}

\usepackage{tikz}
\usetikzlibrary{matrix,arrows}

\usepackage{graphicx}
\usepackage{dcolumn}
\usepackage{bm}

\begin{document}


	\title[Exclusive Ferions] {Generalized Fermi-Dirac Distribution of Exclusive Fermions}

\author{Chung-Ru Lee}
\email{math.crlee@gmail.com}

\author{Chin-Rong Lee}
\affiliation{Department of physics\\
National Chung-Cheng University.}
\email{phycrl@ccu.edu.tw}

\date{\today}
\counterwithout{equation}{section}

\begin{abstract}
A system of exclusive fermions occurs when two fermions of opposite spin are prohibited from occupying the same quantum level. We derive the distribution of exclusive fermions via the employment of the grand canonical ensemble. Salient features of its statistical properties, compared to the free electron gases, include: larger Fermi energy, higher degeneracy pressure, but the same Pauli paramagnetism and Landau diamagnetism. In particular, higher degeneracy pressure leads to an inflation of the Chandrasekhar limit to 1.6 times when applied to white dwarf stars and neutron stars.
\end{abstract}
\date{\today}
\maketitle


\section{Introduction}
\subsection{Grand Partition Function for Fermi-Dirac Distribution}
Consider the statistical distribution of a bunch of non-interacting fermions. A quantum state (labeled $|1\rangle$, with energy $\epsilon_1=\epsilon_1^\uparrow=\epsilon_1^\downarrow$ for simplicity) can be empty, or occupied by one fermion of either spin, or by two fermions of opposite spin. The $N$-particle canonical partition function $Z_N$ \cite{1,2,3} can be written with the sum of all possibilities as
\begin{align}
\begin{split}
&Z_N^{\hat{1}}+Z_{N-1}^{\hat{1}}e^{-\beta\epsilon_1^\uparrow}+Z_{N-1}^{\hat{1}}e^{-\beta\epsilon_1^\downarrow}+Z_{N-2}^{\hat{1}}e^{-\beta(\epsilon_1^\uparrow+\epsilon_1^\downarrow)}\\
=&Z_N^{\hat{1}}+2Z_{N-1}^{\hat{1}}e^{-\beta\epsilon_1}+Z_{N-2}^{\hat{1}}e^{-2\beta\epsilon_1}.
\end{split}\label{eqn:FD1}
\end{align}
The superscript $\hat{1}$ indicates that $Z_N^{\hat{1}}$ is the $N$-particle canonical partition function for which the state $|1\rangle$ is removed from the spectrum.

The grand partition function $\mathcal{Z}_G$ is then
\begin{widetext}
\begin{align}
\mathcal{Z}_G&=\sum_{N=0}^\infty Z_Nz^N\qquad(z=e^{\beta\mu}\text{ is the fugacity})\notag\\
&=\sum_{N=0}^\infty Z_N^{\hat{1}}z^N+2e^{-\beta\epsilon_1}z\sum_{N=0}^\infty Z_{N-1}^{\hat{1}}z^{N-1}+e^{-2\beta\epsilon_1}z^2\sum_{N=0}^\infty Z_{N-2}^{\hat{1}}z^{N-2}\notag\\
& =(1+2e^{-\beta\epsilon_1}z+e^{-2\beta\epsilon_1}z^2)\mathcal{Z}_G^{\hat{1}}\notag\\
& =(1+e^{-\beta\epsilon_1}z)^2\mathcal{Z}_G^{\hat{1}}\notag\\
& =\prod_i(1+e^{-\beta\epsilon_i}z)^2.
\end{align}
\end{widetext}
The last equality is achieved by specifying the outcome for all single-particle states inductively.

The expected value of the number of particles is given by
$$\langle N\rangle=z\frac{\partial\log \mathcal{Z}_G}{\partial z}=\sum_i\frac{2}{e^{\beta(\epsilon_i-\mu)}+1}=\sum_i\langle n_i\rangle.$$
This gives the Fermi-Dirac distribution. The factor 2 here is often regarded as the spin degeneracy and absorbed into the density of states (which doubles the counting of states). We will instead treat it as an intrinsic factor in the distribution function
$$f(\epsilon)_\mathrm{FD}=\frac{2}{e^{\beta(\epsilon-\mu)}+1}.$$

\subsection{Exclusive Fermions}
In certain situations, for instance, in doped semiconductors or electron gases under an extremely high magnetic field, the Coulomb repulsion between electrons may (effectively, through energetic constraint) prevent electrons from double occupancy of a single state, regardless of their spin \cite{4,5,6}. Though in a distinct manner, taking Coulomb blockade into account is also the stream of ideas that led to $t-J$ model, a prospective microscopic theory of high-temperature superconductivity.

We would like to refer to fermions in such systems as \textit{exclusive fermions}, as they present stronger exclusiveness.

The statistical distribution of exclusive fermions can be derived by dropping the double occupancy term in equation (\ref{eqn:FD1}). The grand canonical partition function $\mathcal{Z}_G$ is then
$$\mathcal{Z}_G=(1+2e^{-\beta\epsilon_1}z)\mathcal{Z}_G^{\hat{1}}=\prod_i(1+2e^{-\beta\epsilon_i}z).$$
From this, we easily yield the distribution function of exclusive fermions:
\begin{align}
f(\epsilon)=\frac{2}{e^{\beta(\epsilon-\mu)}+2}.
\end{align}

\section{The Equation of State}
Let us inspect an exclusive fermion gas. For non-relativistic particles of mass $m$, the density of states at a given volume $V$ is
\begin{align}
\mathcal{D}(\epsilon)=\frac{V}{4\pi^2\hbar^3}(2m)^{3/2}\epsilon^{1/2}=:bV\epsilon^{1/2}.
\end{align}
The total number of particles within the volume $V$, under the distribution function above, gives
$$\langle N\rangle=\int_0^\infty\mathcal{D}(\epsilon)\frac{2}{e^{\beta\epsilon}z^{-1}+2}d\epsilon=bV\int_0^\infty\frac{2\epsilon^{1/2}}{e^{\beta\epsilon}z^{-1}+2}d\epsilon.$$
The average energy of the gas is
$$\langle E\rangle=\int_0^{\infty}\mathcal{D}(\epsilon)\frac{2\epsilon}{e^{\beta\epsilon}z^{-1}+2}d\epsilon=bV\int_0^\infty\frac{2\epsilon^{3/2}}{e^{\beta\epsilon}z^{-1}+2}d\epsilon$$
The pressure can be calculated in the grand canonical ensemble using
$$PV=\frac{1}{\beta}\log \mathcal{Z}_G=\frac{1}{\beta}\int_0^\infty\epsilon^{1/2}\log(1+2e^{-\beta\epsilon}z)d\epsilon.$$
Integration by parts, we have
\begin{align}
PV&=-\frac{bV}{\beta}\frac{2}{3}\int_0^\infty\frac{-2\beta\epsilon^{3/2}}{e^{\beta\epsilon}z^{-1}+2}d\epsilon\notag\\
&=\frac{2}{3}\langle E\rangle.
\end{align}
This is the equation of state. At high temperature (when $z\ll 1$), we can keep only the leading correction term. Therefore (apply $x=\beta\epsilon$)
\begin{align*}
\langle N\rangle&=2bV\int_0^\infty\frac{\epsilon^{1/2}}{e^{\beta\epsilon}z^{-1}+2}d\epsilon\\
&=\frac{2bVz}{\beta^{3/2}}\int_0^\infty\frac{x^{1/2}e^{-x}}{1+2e^{-x}z}dx\\
&\approx\frac{2bVz}{\beta^{3/2}}\int_0^\infty{x^{1/2}e^{-x}}(1-2e^{-x}z)dx\\
&=\frac{2bVz}{\beta^{3/2}}(\frac{\sqrt{\pi}}{2}-z\frac{\sqrt{2}}{4}\sqrt{\pi}).
\end{align*}
Likewise,
$$\langle E\rangle\approx\frac{2bVz}{\beta^{5/2}}(\frac{3}{4}{\sqrt{\pi}}-z\frac{3\sqrt{2}}{16}\sqrt{\pi}).$$

These quantities can be simplified using the thermal wavelength $\lambda=\sqrt{\frac{2\pi\hbar^2}{mk_\mathrm{B}T}}$ and the relation $\frac{b}{\beta^{3/2}}\frac{\sqrt{\pi}}{2}=\lambda^{-3}$. We then have
\begin{align}
\begin{cases}
\frac{\langle N\rangle}{V}=\frac{2}{\lambda^3}z(1-\frac{z}{\sqrt{2}})\\
\frac{\langle E\rangle}{V}=\frac{1}{\beta}\frac{3}{\lambda^3}z(1-\frac{z}{2\sqrt{2}})
\end{cases}\label{eq:states}
\end{align}

Here, the high temperature limit means that the thermal wavelength is much less than the interparticle spacing, that is, $\frac{\lambda^3}{V/\langle N\rangle}\ll1$. Then from (\ref{eq:states}), we have
$$z\approx\frac{\lambda^3}{2}\frac{\langle N\rangle}{V}(1+\frac{1}{\sqrt{2}}\frac{\lambda^3}{2}\frac{\langle N\rangle}{V})$$
Again by (\ref{eq:states}),
$$\langle E\rangle\approx\frac{3}{2}\langle N\rangle k_\mathrm{B}T(1+\frac{1}{4\sqrt{2}}\lambda^3\frac{\langle N\rangle}{V}).$$
Therefore, at high temperatures, the ideal gas of exclusive fermions satisfies
\begin{align}
PV\approx\langle N\rangle k_\mathrm{B}T(1+\frac{1}{4\sqrt{2}}\lambda^3\frac{\langle N\rangle}{V}).
\end{align}

In comparison to the classical ideal gas, the quantum correction increases the pressure. This is the degeneracy pressure caused by the fermionic characteristics. We would like to remark that the degeneracy pressure calculated here is greater than that of the usual Fermi particles under the Pauli Exclusion Principle.

\section{The Fermi Surface}
At absolute zero $T=0$, the distribution function $f(\epsilon)$ becomes a step function, which means that a state is either filled or empty, depending on whether $\epsilon\leq\mu(0)$. Here, let $\mu(0)$ be the chemical potential at $T=0$, it is called the Fermi energy. We will write $\mu(0)=:E_F$ \cite{7,8}.

For $N$ particles in a box with volume $V$, we have
$$N=\int_0^\infty\mathcal{D}(\epsilon)f(\epsilon)d\epsilon.$$
Take $T=0$, we yield
$$N=\int_0^{E_F}\mathcal{D}(\epsilon)d\epsilon=\frac{2}{3}bVE_F^{3/2}.$$
Or equivalently,
$$E_F=\frac{\hbar^2}{2m}(6\pi^2\frac{N}{V})^{2/3}.$$
Note that the Fermi energy of exclusive fermions turns out to be $2^{2/3}\approx1.6$ times bigger than that of the free electron gases.

The total energy is related to the Fermi energy in a simple way:
\begin{align}
E=\int_0^{E_F}\epsilon\mathcal{D}(\epsilon)d\epsilon=bVE_F^{5/2}=\frac{3}{5}NE_F.
\end{align}

An immediate application of this result is to consider the white dwarf stars or neutron stars. When a star exhausts its fuel, it has to rely on degeneracy pressure to resist gravitational force. When the star has impurities (that behave like doped semiconductors) or possesses a strong magnetic field (like magnetars), the fermions (electrons or neutrons) may become exclusive. In such cases, they obey the distribution law for exclusive fermions given by $f(\epsilon)$. The degeneracy pressure given by
$$P_\text{deg}=-\frac{\partial E}{\partial V}=\frac{2}{5}NE_F/V$$
is proportional to the Fermi energy $E_F$ when the particle density $N/V$ is constant. Following standard calculations, the stronger degeneracy pressure would withstand a larger (in fact, about $1.6$ times) estimation of the Chandrasekhar limit \cite{9,10}.

\section{Heat Capacity at Low Temperature}
At $T>0$, we consider integrals of the form
\begin{align}
J=-\int_0^\infty g(\epsilon)\frac{\partial f}{\partial\epsilon}d\epsilon.
\end{align}

We follow the method of Sommerfeld expansion. At low temperature $(k_\mathrm{B}T\ll\mu)$, we may expand $g(\epsilon)$ around $\epsilon=\mu$ to yield
\begin{align*}
J=&-g(\mu)f|_{\epsilon=0}^\infty-g'(\mu)\int_0^\infty(\epsilon-\mu)\frac{\partial f}{\partial\epsilon}d\epsilon\\
&-\frac{1}{2}g''(\mu)\int_0^\infty(\epsilon-\mu)^2\frac{\partial f}{\partial\epsilon}d\epsilon+\dots.
\end{align*}
Note that $f\rightarrow0$ as $\epsilon\rightarrow\infty$, while at low temperature, $f(0)\approx1$. Therefore, we can replace the leading term by a simple $g(\mu)$. Let $x=\frac{\epsilon-\mu}{k_\mathrm{B}T}$, then (at low temperature), $J$ is approximated by
\begin{align*}
&g(\mu)+g'(\mu)k_\mathrm{B}T\int_{-\infty}^\infty\frac{2xe^x}{(e^x+2)^2}dx\\
+&\frac{1}{2}g''(\mu)(k_\mathrm{B}T)^2\int_{-\infty}^\infty\frac{2x^2e^x}{(e^x+2)^2}dx.
\end{align*}
For brevity, we will set
\begin{align}
A_n&=\int_{-\infty}^\infty\frac{x^ne^x}{(e^x+2)^2}dx.
\end{align}
Thus
\begin{align}
J\approx g(\mu)+2g'(\mu)k_\mathrm{B}TA_1+g''(\mu)(k_\mathrm{B}T)^2A_2.
\end{align}
Numerically, $A_1\approx0.34657$ and $A_2\approx1.88516$. Note that in contrast to the computation in, say, free electron gases, the formula here involves $g'(\mu)$ due to the asymmetry in $f(\epsilon)$.

Integration by parts gives
$$N=-\frac{2}{3}bV\int_0^\infty\epsilon^{3/2}\frac{\partial f}{\partial\epsilon}d\epsilon$$
and
$$E=-\frac{2}{5}bV\int_0^\infty\epsilon^{5/2}\frac{\partial f}{\partial\epsilon}d\epsilon$$
are both written in the form of $J$ and can be estimated by the approximation above. In particular,

\begin{align}
N&\approx\frac{2}{3}bV\left(\mu^{\frac{3}{2}}+3\mu^{\frac{1}{2}}k_\mathrm{B}TA_1+\frac{3}{4}\mu^{-\frac{1}{2}}(k_\mathrm{B}T)^2A_2\right).\notag\\
& =\frac{2}{3}bV\mu^{3/2}\left(1+3\frac{k_\mathrm{B}T}{\mu}A_1+\frac{3}{4}(\frac{k_\mathrm{B}T}{\mu})^2A_2\right)
\label{eq:heat1}\\
E&\approx\frac{2}{5}bV\mu^{3/2}\left(1+5\frac{k_\mathrm{B}T}{\mu}A_1+\frac{15}{4}(\frac{k_\mathrm{B}T}{\mu})^2A_2\right)\label{eq:heat2}
\end{align}
Since the total number of particles is conserved for the change of temperature, we compare the result in (\ref{eq:heat1}) to the number at $T=0$:
$$\frac{2}{3}bVE_F^{3/2}=N.$$
This leads to
$$E_F^{3/2}\approx\mu^{3/2}\left(1+3A_1\frac{k_\mathrm{B}T}{\mu}+\frac{3}{4}A_2(\frac{k_\mathrm{B}T}{\mu})^2\right)$$
and so we can express the chemical potential at finite temperature in terms of the Fermi energy:
$$\mu\approx E_F\left(1-2A_1\frac{k_\mathrm{B}T}{E_F}-(\frac{A_2}{2}-9A_1^2)(\frac{k_\mathrm{B}T}{E_F})^2\right).$$
As expected, the chemical potential is at its maximum when $T=0$ and decreases as the temperature rises.

When considering the quotient of (\ref{eq:heat1}) and (\ref{eq:heat2}), we get
$$\frac{E}{N}\approx\frac{3}{5}(\mu+2A_1k_\mathrm{B}T+\mu^{-1}(3A_2-6A_1^2)(k_\mathrm{B}T)^2).$$
And so the specific heat per particle becomes
\begin{align}
c=\frac{\partial}{\partial T}(\frac{E}{N})\approx(3A_2-\frac{6}{5}A_1^2)\frac{k_\mathrm{B}^2}{E_F}T\approx5.55\frac{k_\mathrm{B}^2}{E_F}T.
\end{align}
Under the same Fermi energy, the specific heat for such a system is slightly greater than the $4.93\frac{k_\mathrm{B}^2}{E_F}T$ for fermion gases.

\section{Pauli paramagnetism}
We now consider a gas of exclusive electrons in a constant magnetic field $B$. Under the field $B$ (of strength $B$ by an abuse of notation), the kinetic energy of an electron picks up an extra term
$$E_\text{spin}=\pm\mu_\mathrm{B}B,$$
where $\mu_\mathrm{B}=\frac{e\hbar}{2m_\mathrm{e}}$ is the Bohr magneton, and the sign depends on the spin of the electron.

The number density of up-spinning electrons is given by
\begin{align*}
\frac{N_\uparrow}{V}&=b\int_0^\infty\frac{\epsilon^{1/2}}{z^{-1}e^{\beta(\epsilon+\mu_\mathrm{B}B)}+2}d\epsilon\\
&=b\frac{y}{\beta^{3/2}}\int_0^\infty\frac{x^{1/2}e^{-x}}{1+2ye^{-x}}dx\\
&=b\frac{y}{\beta^{3/2}}\int_0^\infty x^{1/2}e^{-x}(1-2ye^{-x}+\dots)dx.
\end{align*}
Here, we made the substitution $y=ze^{-\beta\mu_\mathrm{B}B}$ and $x=\beta\epsilon$. The leading term gives
$$N_\uparrow\approx\frac{V}{\lambda^3}ze^{-\beta\mu_\mathrm{B}B}.$$
Analogously, from the number density of down-spinning electrons we have
$$N_\downarrow\approx\frac{V}{\lambda^3}ze^{\beta\mu_\mathrm{B}B}.$$
The magnetization $M$ depends on the difference between $N_\uparrow$ and $N_\downarrow$:
$$M=-\mu_\mathrm{B}(N_\uparrow-N_\downarrow)\approx\frac{V}{\lambda^3}\cdot2\mu_\mathrm{B}z\sinh(\beta\mu_\mathrm{B}B).$$
Since the total number is simply the sum
\begin{align}
N=N_\uparrow+N_\downarrow\approx\frac{V}{\lambda^3}\cdot2z\cosh(\beta\mu_\mathrm{B}B),
\end{align}
we obtain from this the high temperature magnetization of exclusive electrons:
$$M\approx N\mu_\mathrm{B}\tanh(\beta\mu_\mathrm{B}B).$$
This turns out to be the same as the Pauli paramagnetization.

\section{Landau Diamagnetism}
Landau diamagnetism arises from the quantized orbital motion of the electrons in a magnetic field. From the spectral point of view, it forms the well-known Landau levels. Let us compute the grand partition function for exclusive electrons in a box of volume $V=L^3$, with the presence of a constant magnetic field of strength $B$ along the $z$-direction.

We have
\begin{align}
\log \mathcal{Z}_G=\left(\frac{gL}{2\pi\hbar}\right)\sum_{n=0}^\infty\int_{-\infty}^\infty\log(1+2ze^{-\beta\epsilon(p,n)})dp.
\end{align}
Here $g$ is the degeneracy of a Landau level, given by $g=\frac{eBL^2}{2\pi\hbar}$, $p=p_z$ is the momentum in the $z$-direction, $n$ is the index for the Landau levels, and the energy is $\epsilon=\frac{p^2}{2m}+\hbar\omega_\mathrm{c}(n+\frac{1}{2})$ with cyclotron frequency $\omega_\mathrm{c}=\frac{eB}{m}$.

We consider the magnetization in the classical domain. Therefore, we take the high-temperature limit. That is, to perform an approximation where $z$ is small, and retain the first-order terms:
\begin{align*}
\log \mathcal{Z}_G & \approx2\left(\frac{gL}{2\pi\hbar}\right)\sum_{n=0}^\infty\int_0^\infty2ze^{-\beta\left(\frac{p^2}{2m}+\hbar\omega_\mathrm{c}(n+\frac{1}{2})\right)}dp\\
&=\frac{2zgL}{\pi\hbar}\sum_{n=0}^\infty e^{-\beta\hbar\omega_\mathrm{c}(n+\frac{1}{2})}\int_0^\infty e^{-\beta\frac{p^2}{2m}}dp\\
&=\frac{2zgL}{\pi\hbar}\frac{e^{-s}}{1-e^{-2s}}\left(\frac{1}{2}\sqrt{\frac{2\pi m}{\beta}}\right)\\
&\approx\frac{2zgL}{\pi\hbar}\frac{1}{2s}\left(1-\frac{s^2}{6}\right)\left(\frac{1}{2}\sqrt{\frac{2\pi m}{\beta}}\right)\\
&=\frac{2zV}{\lambda^3}\left(1-\frac{1}{2\lambda}\left(\frac{\hbar\omega_\mathrm{c}}{k_\mathrm{B}T}^2\right)\right),
\end{align*}
where the thermal wavelength $\lambda=\sqrt{\frac{2\pi\hbar^2}{mk_\mathrm{B}T}}$, and we let $s=\frac{\hbar\omega_\mathrm{c}}{2k_\mathrm{B}T}$ in the computation.

To eliminate the fugacity $z$, we use its relation to the particle number $N$:
\begin{align*}
N&=\frac{gL}{2\pi\hbar}\sum_{n=0}^\infty\int_{-\infty}^\infty\frac{2}{z^{-1}e^{\beta\epsilon}+2}\\
&\approx 2\frac{zV}{\lambda^3}
\end{align*}
as taking the first-order approximation to what was computed in (\ref{eq:states}).

As a conclusion, we obtain the magnetic susceptibility
\begin{align}
\chi_\mathrm{dia}\approx-\frac{1}{3k_\mathrm{B}T}\frac{N}{V}\left(\frac{e\hbar}{2m}\right)^2=-\frac{1}{3}\frac{\mu_\mathrm{B}^2}{k_\mathrm{B}T}\left(\frac{N}{V}\right),
\end{align}
same as the result of Landau for free electron gases.


\bibliography{refs}{}

\end{document}